# Synthesis and properties of a novel narrow band gap oligomeric diketopyrrolopyrrole-based organic semiconductor

Mylène Le Borgne[a,b], Jesse Quinn[b], Jaime Perez[c], Natalie Stingelin[c], Guillaume Wantz[a]* and Yuning Li[b]*

An oligomeric semiconductor containing three bisthiophenediketopyrrolopyrole units (Tri-BTDPP) was synthesized and characterized. Tri-BTDPP has a HOMO level of -5.34 eV, a broad absorption close to the near infrared region and a low band gap of 1.33 eV. Additionally, a promising hole mobility of 1 x $10^{-3}$ cm² $V^{-1}$ $s^{-1}$ was achieved after thermal annealing at 150 °C in organic field effect transistors (OFET). Organic photovoltaic (OPV) cells containing Tri-BTDPP and $PC_{71}BM$ as the donor/acceptor couple exhibited a power conversion efficiency (PCE) of 0.72%. Through an intensive study of the active layer using AFM, XRD, and DSC, it was found that Tri-BTDPP and $PC_{71}BM$ were unable to intermix effectively, resulting in oversized Tri-BTDPP crystalline phases and thus poor charge separation. Strategies to improve the OPV performance were thus proposed.

[a.] *Univ. Bordeaux, IMS, UMR 5218, F-33400, Talence, France ;
CNRS, IMS, UMR 5218, F-33400, Talence, France
Bordeaux INP, IMS, UMR 5218, F-33400, Talence, France*
[b.] *Dept of Chemical Engineering, University of Waterloo, 200 University Ave West, Waterloo, ON N2L 3G1, Canada*
[c.] *Dept of Materials and Centre for Plastic Electronics, Imperial College London, London, UK*

# 1. Introduction

Organic photovoltaic (OPV) solar cells represent a promising technology to provide sustainable and cost-effective energy, which can be produced using printing technologies. They have attracted enormous attention from the scientific community and industry for the last 20 years. Semiconducting polymers were the first promising materials to be introduced in OPV devices. Recently, however solution-processed small molecules [1–4] have demonstrated very promising performance and are currently receiving increasing attention. Small molecule organic semiconductors have well-defined structures and much higher purity than their polymer counterparts. Their syntheses do not suffer from large batch-to-batch variation as those of polymers, although small molecules present poor film formation capability by solution processing compared with polymers. Consequently, OPVs based on solution-processed small molecules often suffer from low fill factors (FF). To improve the solution processability while maintaining other merits of small molecules, oligomeric organic semiconductors with molecular weights between those of polymers and small molecules are being studied extensively. The best power conversion efficiencies (PCE) to date are in the range of 6-9%,[5–7] values comparable to those of polymer solar cells.[8–10]

Diketopyrrolopyrrole (DPP) is currently the most widely studied electron-accepting building block for constructing low band gap semiconductors including small molecules, oligomers, and polymers for OPVs. DPP-based solution-processed small molecules were first introduced by Nguyen and co-workers in 2008 as donors in OPVs.[11] Since then, various solution processed small molecules/oligomers based on DPP have been reported with PCEs as high as 5%.[12–15] A series of oligomers consisting of two DPP terminal units and different cores such as naphtodithiophene and benzodithiophene (DPP-Core-DPP) were reported to achieve a PCE of 4-5%.[15,16] In 2013, Nguyen's group reported two tri-DPP (DPP trimers) made of one bisphenyl-DPP and two bisthiophene DPP.[12,17] The use of tri-DPP improved the charge transport, the quality of the film and the fill factor of the device. These improvements resulted in a PCE of 5.5%. Here, we report a Tri-DPP oligomer consisting of three bisthiophene DPP units in order to achieve a coplanar structure. A coplanar molecular geometry is expected to contribute to greater crystallinity, improved charge transport, and a lower band gap of the semiconductor. We used this new DPP compound as an electron donor (combined with $PC_{71}BM$) in OPV devices.

# 2. Experimental section

## 2.1 Materials and measurements

All reagents and chemicals were purchased from commercial sources (Sigma Aldrich, Alpha Aesar) and used without further purification. 3-(5-Bromothiophen-2-yl)-2,5-bis(2-ethylhexyl)-6-(thiophen-2-yl)-2,5-dihydropyrrolo[3,4-c]pyrrole-1,4-dione (1) and 2,5-bis(2-octyldodecyl)-3,6-bis(5-(trimethylstannyl)thiophen-2-yl)-2,5-dihydropyrrolo[3,4-c]pyrrole-1,4-dione (2) were prepared as described in the literature.[18,19]

All $^1$H NMR and $^{13}$C NMR spectra were recorded on a 400 MHz and 300 MHz Bruker NMR using $CDCl_3$ as solvent and tetramethylsilane (TMS, 0 ppm) as a reference. High resolution electrospray technique mass spectrometry was used for Tri-BTDPP and MALDI-TOF for by-product. Cyclic voltammetry (CV) measurements were conducted using a DY2000EN electrochemical workstation in a 0.1 M tetrabutylammonium hexafluorophosphate acetonitrile electrolyte at room temperature at a scan rate of 50 mV s$^{-1}$. The working electrode and counter electrode were platinum electrodes and the reference electrode was Ag/AgCl (0.1 M) electrode. The reference electrode was calibrated against the redox potential of ferrocene/ferrocenium (Fc/Fc$^+$). The thin film on the Pt disk, formed by drop-casting a chloroform solution was used as the working electrode. AFM imaging was carried out at room temperature using an AFM Nanoman from Bruker Instrument with Nanoscope 5 controller. Images were obtained in tapping mode using silicon tips (PointProbe® Plus AFM-probe, Nanosensors, Switzerland). ,DSC, XRD.

## 2.2 Device and characterization

The photovoltaic properties were tested in conventional solar cells and inverted solar cells following respectively those structures: ITO/PEDOT:PSS/Tri:BTDPP:PC$_{71}$BM/(Ca)/Al and ITO/ZnO$_x$/TriBTDPP:PC$_{71}$BM/MoO$_3$/Ag. For conventional solar cells fabrication, ITO–coated glass substrates were successively cleaned in ultrasonic baths of acetone/ethanol/isopropanol for 10 min, followed by oxygen plasma treatment for 15 min. A water solution of PEDOT:PSS, previously filtrated on 0.2 μm, was deposited by spin-coating at 4000 rpm for 60 sec. The layer was dried in the oven at 100 °C under vacuum for 30 min. Then, a solution of Tri-BTDPP:PC$_{71}$BM (1:1, 15 mg ml$^{-1}$) in chloroform and 1 vol% DIO were spin-coated on top of PEDOT:PSS under nitrogen atmosphere. The thickness of the active layer was 90 nm. The solvent were removed by annealing the active layer at 80 °C for 10 minutes. Calcium (10 nm) and then aluminum (80 nm) were thermally evaporated onto the active layer through shadow masks under 2-4 x 10$^{-6}$ mbar. The effective area was 10 mm².

For inverted solar cell, after similarly pre-cleaned the ITO-coated substrates, a solution of ZnO$_x$ prepared with 0.15 M of zinc acetate and 0.15 M of ethanol amine in ethanol was spin-coated at 2000 rpm for 60 sec. The layer was annealed at 180 °C for 1 h. The active layer was prepared and deposited as for conventional solar cell. To complete the device, molybdenum (IV) oxide (10 nm) and silver (80 nm) were thermally evaporated onto the active layer through shadow masks under 2-4 x 10$^{-6}$ mbar. The effective area was 10 mm².

The devices were characterized using a K.H.S SolarCelltest-575 solar simulator with AM 1.5G filters set at 100 mW cm$^{-2}$ with a calibrated radiometer (IL 1400BL). The current density-voltage (J-V) curves measurements were processed with Labview controlled Keithley 2400 SMU. Devices were characterized under nitrogen in a set of glove boxes ($O_2$ and $H_2O$ < 0.1 ppm).

Bottom gate - bottom contacts field effect transistors were fabricated (OFETs) using *Fraunhofer IPMS* templates of doped Si with 200 nm-thick silicon oxide and gold electrodes. Substrates were cleaned with successive acetone and IPA baths followed by 15 min of UV-ozone treatment. A solution of 5 mg ml$^{-1}$ of Tri-BTDPP in chloroform was spin-coated at 2500 rpm for 40 sec on top of transistor substrates in the nitrogen-filled glovebox to form a 40 nm-thick layer. The dimensions of the channel were L= 5 μm and W= 1 cm. Transistors were measured using an analyzer for semiconductor (Keithley 4200) coupled with a three tip station.

### 2.3 Synthesis

**6,6'-((2,5-bis(2-octyldodecyl)-3,6-dioxo-2,3,5,6-tetrahydropyrrolo[3,4-c]pyrrole-1,4-diyl)bis([2,2'-bisthiophene]-5',5-diyl))bis(2,5-bis(2-ethylhexyl)-3-(thiophen-2-yl)-2,5-dihydropyrrolo[3,4-c]pyrrole-1,4-dione) (Tri-BTDPP).** A 100 ml dry three-necked round bottom flask was charged with compound **1** (1.5 g, 0.0025 mol, 2.1 eq), compound **2** (1.40 g, 0.0012 mol, 1 eq), P(o-tolyl)$_3$ (28.7 mg, 0.09 mmol, 0.08 eq) and anhydrous toluene (49 mL) under argon. Then, Pd$_2$(dba)$_3$ (21.9 mg, 0.024 mmol, 0.02 eq) dissolved in anhydrous toluene (1 mL) was added into the flask through a syringe. The mixture was heated at 120 °C for 60 h during which the color of the solution turned from purple to blue. After cooling down to room temperature, the solution was poured into methanol (600 mL) and stirred for 30 min. The resulting precipitate was filtered and purified using column chromatography on silica gel with a mixture of dichloromethane (DCM) and hexane (with a volume ratio of 6:4) and then pure chloroform as eluent to afford 1.65 g (72%) of the crude product and 0.60 g (26%) of 6,6'-([2,2'-bithiophene]-5,5'-diyl)bis(2,5-bis(2-ethylhexyl)-3-(thiophen-2-yl)-2,5-dihydropyrrolo[3,4-c]pyrrole-1,4-dione) (**3**), a by-product formed by the mono-coupling of compound 1 and 2. The crude product obtained by the first column separation contains the target Tri-BTDPP and the by-product 3-(5'-(2,5-bis(2-ethylhexyl)-3,6-dioxo-4-(thiophen-2-yl)-2,3,5,6-tetrahydropyrrolo[3,4-c]pyrrol-1-yl)-[2,2'-bithiophen]-5-yl)-2,5-bis(2-octyldodecyl)-6-(thiophen-2-yl)-2,5-dihydropyrrolo[3,4-c]pyrrole-1,4-dione (**4**), formed by the homo-coupling of 1. 0.50 g of this crude product was further purified on silica gel column chromatography at 50 °C using a mixture of chloroform/toluene (volume ratio: 1/1) as eluent to afford 0.39 g pure Tri-BTDPP as a dark blue solid and 0.11 g of the by-product **4**. The total amounts of Tri-BTDPP and **4** in 1.65 g of the crude product are estimated to be 1.29 g (56%) and 0.36 g (16%), respectively. Data for Tri-BTDPP: $^1$H NMR (400 MHz; CDCl$_3$) δ 8.93 (2H, d, J = 4.2 Hz), 8.91 (2H, d, J = 4.2 Hz), 8.85 (2H, dd, J = 3.9, 1.1 Hz), 7.49 (2H, dd, J = 5.0, 1.0 Hz), 7.25 (4H, dd, J = 3.9, 3.5 Hz), 7.14 (2H, dd, J = 4.9, 4.0 Hz), 3.96 (12H, d, J=5.2 Hz), 2.00 - 1.71 (6H, m), 1.39 - 1.23 (35H, m), 1.22 - 1.05 (70H, m), 0.92 – 0.72 (41H, m) .$^{13}$C NMR (400 MHz, CDCl$_3$) δ 161.81, 161.69, 141.49, 141.19, 140.67, 139.33, 139.28, 137.43, 137.15, 135.95, 131.03, 130.14, 130.02, 129.95, 128.77, 126.12, 109.37, 109.08, 108.44, 77.56, 46.76, 46.30, 39.73, 39.46, 38.41, 32.26, 32.24, 31.60, 30.68, 30.56, 30.51, 30.45, 30.05, 29.99, 29.97, 29.92, 29.72, 29.70, 28.84, 28.69, 26.65, 24.02, 23.87, 23.47, 23.44, 23.03, 14.45, 14.38, 10.91, 10.83. HR-MS: m/z 1906.11094 (M$^+$, 90%)

Data for **3**: NMR (300 MHz; CDCl$_3$) δ 9.05-8.84 (4H, m), 7.65 (2H, d, J=4.9 Hz), 7.43 (2H, d, J=4.2 Hz), 7.33 – 7.28 (1H, m), 7.23 – 7.18 (1H, m), 4.04 (8H, d, J= 7.1 Hz), 1.92 (4H,s), 1.44 – 1.07 (8H, m), 1.04-0.67 (26H, m).

Data for **4**: MALDI-TOF: m/z 1046.4 (M$^+$, 100%)

## 3. Results and discussion

### 3.1 Synthesis and characterization

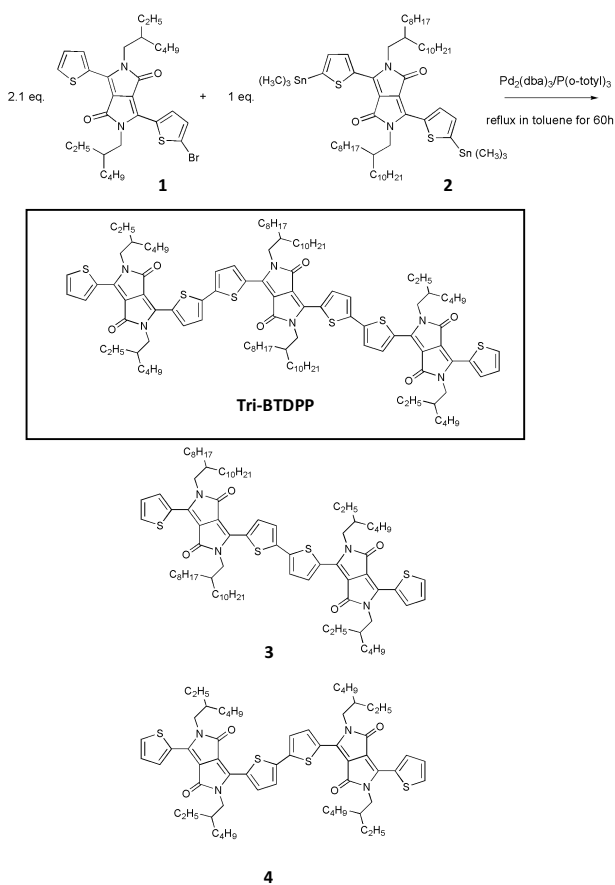

**Scheme 1** Synthetic route to Tri-BTDPP via Stille coupling reaction of 1 and 2.

The synthesis route to 6,6'-((2,5-octyldodecyl)-3,6-dioxo-2,3,5,6-tetrahedropyrrolo[3,4-c]pyrrole-1,4-diyl)bis([2,2'-bisthiophene]-5',5-diyl))bis(2,5-bis(2-ethylhexyl)-3-(thiophen-2-yl)-2,5-dihydropyrrolo[3,4-c]pyrrole-1,4-dione) (Tri-BTDPP) is shown in Scheme 1. The mono-brominated DPP compound **1** and the bis(trimethylstannyl) DPP compound **2** were synthesized according to the methods reported in the literature[18,19]. The introduction of the long 2-octyldodecyl side chain on **2** was intended to render the final compound Tri-BTDPP solution-processable. The synthesis of Tri-BTDPP was conducted via Stille cross coupling reaction between **1** and **2** in a 2.1:1 molar ratio in toluene under reflux for 60 h in the presence of $Pd_2(dba)_3$ / $P(o\text{-tolyl})_3$ as a catalyst. Separation of the target product Tri-BTDPP was accomplished by column chromatography, which was not straightforward due to the close polarity of the Tri-BTDPP and a by-product **4**, which was formed by the homo-coupling of compound **1**. First, column chromatography on silica gel using DCM:hexane and then chloroform was conducted to remove compound **3** (26%) and other impurities to yield a crude product (72%), which still contained compound 4. Further purification was conducted by using a heated (50 °C) silica gel column using a mixture of chloroform and toluene (with a volume ratio of 1:1) as eluent, which enabled separation of Tri-BTDPP and **4**. Pure Tri-BTDPP was obtained in 56% yield, with the molecular structure confirmed by NMR and ESI mass spectrometry. The yield of homo-coupling dimer by-product **4** yield was 16%. Compound 4 was characterized by MALDI-TOF mass spectrometry. NMR spectra could not be obtained for **4** due to the strong aggregation tendency of this compound. Formation of a notable quantity (~16%) of homo-coupling by- product in a Stille coupling of a similar brominated DPP compound under similar conditions was recently reported.[20] Other types of catalysts such as $Pd(PPh_3)_4$ may be used to improve the yields of the target Tri-BTDPP.

Among the solvents tested (hexane, chloroform, toluene, chlorobenzene and o-dichlorobenzene), only chloroform provides a solubility above 10 mg mL$^{-1}$ for Tri-BTDPP, which confirms the need of long alkyl chains on the central DPP unit. The thermal properties of Tri-BTDPP were studied by differential scanning calorimetry (DSC). Tri-BTDPP presents an exothermic peak corresponding to its melting point at 253 °C in the heating scan and an endothermic peak at 232 °C corresponding to its crystallization temperature in the cooling scan. Tri-BTDPP has a strong tendency to crystallize as its crystallization enthalpy is equal to 87% of its melting enthalpy.

### 3.2 Optical properties

The UV-Vis absorption spectra of Tri-BTDPP in chloroform solution and neat film are shown in Fig. 1. Tri-BTDPP absorbs in the range of 600 to 850 nm in solution with the maximum absorption peak ($\lambda_{max}$) at 707 nm ($\lambda_{max}$). Compared with the Tri-DPP containing a central bisphenyl DPP unit, which shows absorption onsets at ~670 nm,[12,17] Tri-BTDPP presents a much longer absorption onset at ~810 nm, owing to the increased conjugation length as a result of higher coplanarity and a more efficient intramolecular charge transfer from the electron-donating thiophene units to the central electron-accepting DPP unit. In film, the $\lambda_{max}$ red-shifts to 725 nm and an additional shoulder at ~830 nm appears. The spectrum extends up to 1000 nm. The red-shift of the absorption spectrum from solution to film originates from the planarization of the molecule and the strong intermolecular interactions in the solid state. The optical band gap ($E_g$) was calculated from the onset absorption wavelength to be 1.33 eV, which fell in the range of 1.2 eV to 1.7 eV needed for optimal harvesting of sunlight to achieve a high PCE.[22]

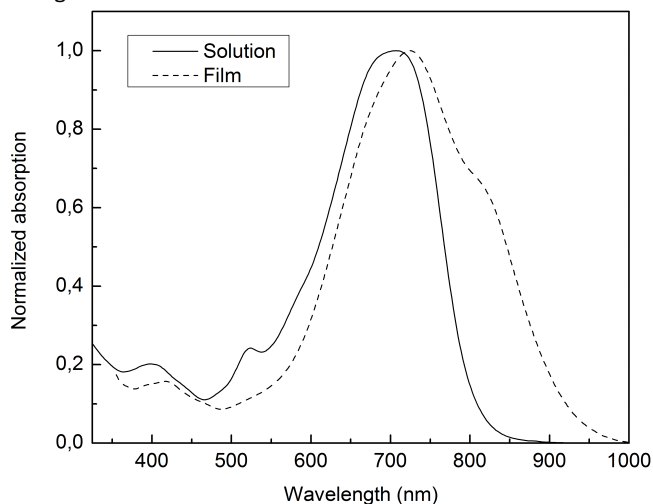

**Fig. 1** UV-Visible spectra of tri-BTDPP in solution and in film.

### 3.3 Electrochemical properties and theoretical calculations

Density functional theory (DFT) calculations were carried out on Tri-BTDPP to study its optimizedmolecular geometry and energy levels. The Beckès three parameter exchange functional and LYP correlation functional (B3LYP) were employed on the standard 6-31G basis set to define the lowest energy conformation. In addition, methyl side chains were chosen to speed up the calculation. As shown in Fig. 2, Tri-BTDPP assumes a coplanar structure as the lowest energy conformation, which would be beneficial for intermolecular interactions and close π-π stacking. It can also be seen that the highest occupied molecular orbital (HOMO) and the lowest unoccupied molecular orbital (LUMO) extend over the entire molecule, revealing an efficient conjugation and intramolecular charge transfer (ICT) between thiophene and DPP moieties. The theoretical values of the HOMO and LUMO levels were evaluated to be -4.76 eV and -3.15 eV, respectively.

The electrochemical properties of Tri-BTDPP were investigated by cyclic voltammetry (CV). The data are summarized in Table 1. The HOMO level was calculated from the onset of the first oxidation peak using the equation: $E_{HOMO}$ = $-(E_{ox}^{onset} – (E^{Fc}-4.8))$ eV, where -4.8 eV is the HOMO energy level of ferrocene (Fc).[21] Since no reduction peak was observed, the LUMO was estimated by adding the optical band gap to the HOMO. The HOMO and the LUMO of Tri-BTDPP were respectively evaluated to be -5.34 eV and -3.99 eV. For a given acceptor, the HOMO level of a donor dictates the open circuit voltage ($V_{oc}$) of the resulting OPV devices. Based on Scharber's methods,[22] a theoretical $V_{oc}$ value of 0.74 eV can be attributed for a blend of Tri-BTDPP and PCBM. Additionally, they reword a diagram predicting the best PCE that can be achieved by a specific donor/PCBM system according to the energy levels. Using Tri-BTDPP as donor material, a PCE of 10% can be predicted. Although achieving such a PCE requires an ideal morphology and efficient and balanced charge transport, this analysis confirmed that Tri-BTDPP is a promising donor for organic solar cells.

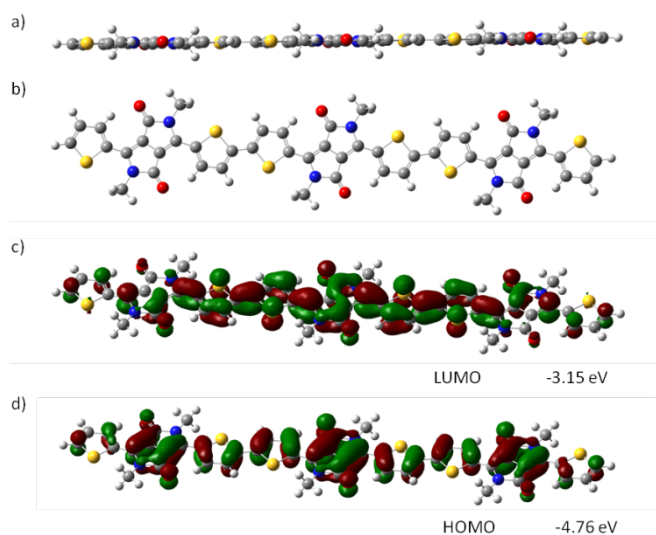

**Fig. 2** DFT calculations of Tri-BTDPP. a-b) Lowest energy conformation; c-d) Frontier orbitals.

**Table 1** Theoretical and experimental energy levels.

|  | HOMO (eV) | LUMO (eV) | $E_g$ (eV) |
| --- | --- | --- | --- |
| **DFT** | -4.76 | -3.15 | 1.61 |
| **CV** | -5.34 | -3.99 | 1.33 |

### 3.4 Charge transport and photovoltaic properties

The charge transport property of Tri-BTDPP was studied using bottom-gate bottom-contact field effect transistors (OFETs) using n-doped $Si/SiO_2$ wafer substrates. The devices showed p-type hole transport performance. The average saturation hole mobility of the as-cast thin films was evaluated to be $1.21\times10^{-4}$ $cm^2$ $V^{-1}$ $s^{-1}$. The devices with Tri-BTDPP films annealed at 100 °C and 150 °C for 10 min exhibited higher hole mobilities with an average value of $7.44 \times 10^{-4}$ $cm^2$ $V^{-1}$ $s^{-1}$ and $9.41 \times 10^{-4}$ $cm^2$ $V^{-1}$ $s^{-1}$ respectively, indicating improved molecular ordering in the films upon annealing. Fig. 3 shows the transfer and output curves of the best device, annealed at 150 °C, which shows the highest mobility of $1.22 \times 10^{-3}$ $cm^2$ $V^{-1}$ $s^{-1}$.

The photovoltaic properties of a blend of Tri-BTDPP as a donor and [6,6]-phenyl-$C_{71}$- butyric acid methyl ester ($PC_{71}BM$) as an acceptor were investigated in bulk heterojunction (BHJ) solar cells. Here, we detail several optimizations made of the device architectures and the ink formulation. First, a conventional architecture, ITO/PEDOT:PSS/Tri-BTDPP:$PC_{71}BM$/Al, was used. The active layer was deposited by spin-coating a solution of Tri-BTDPP:$PC_{71}BM$ in chloroform. After optimizing the thickness of the active layer and the Tri-BtDPP:$PC_{71}BM$ ratio, the PCE was < 0.1% ($V_{oc}$= 0.36 V; (FF)= 0.28; short circuit current density ($J_{sc}$) = 0.56 mA $cm^{-2}$). The poor performance was most likely due to a poor morphology, specifically, a poor nanophase separation of the donor and acceptor in the films (see discussion below). Using solvent additives is an efficient way to tailor the morphology and to improve the solar cell performances. [23,24] We used 1,8-diiodooctane (DIO), which is one of the most commonly used additives, to improve the film morphology of the Tri-BTDPP:$PC_{71}BM$ blend. Different concentrations (vol% relative to the solvent) of DIO (0.25%, 0.5%, 1%, 2%, 3%, 4% and 5%) were added in the blends. With 1% DIO, a PCE of 0.27% was obtained. Further increase in DIO concentration did not result in further improvement in the cell performance.

In order to improve the charge collection, a thin layer (~10 nm) of calcium (Ca) was introduced between the active layer and the aluminum electrode. The PCE increased to 0.43%. Next, an inverted solar cell structure, ITO/$ZnO_x$/Tri-BTDPP:$PC_{71}BM$/$MoO_3$/Ag, was adopted to further improve the charge collection. Higher $J_{sc}$ of 2.15 mA $cm^{-2}$ and $V_{oc}$ of 0.67 V were achieved, which led to an increased PCE up to 0.72%. However, the cell performance is still very low compared to other small molecules reported in the literature, mainly due to the low $J_{sc}$.[25]

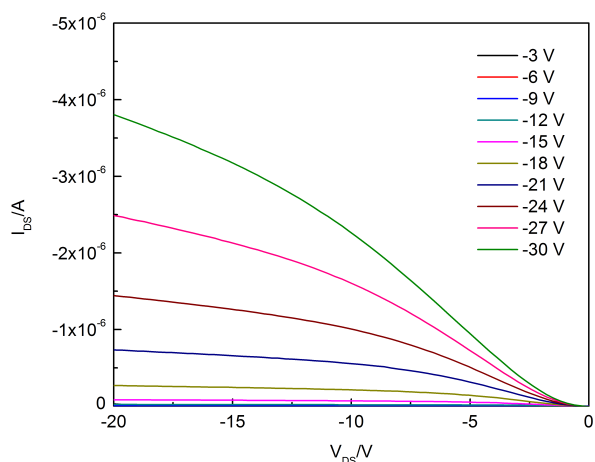

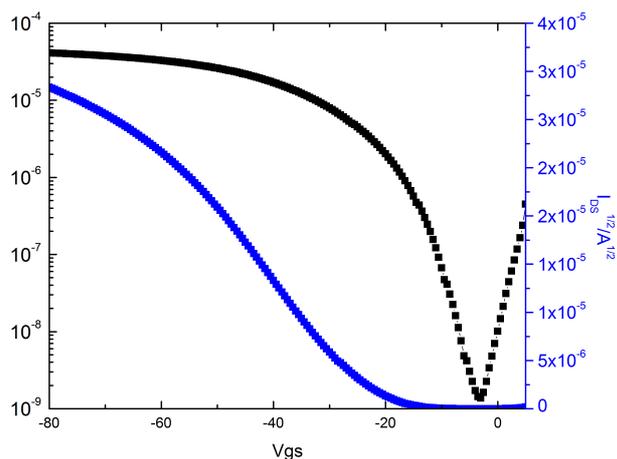

**3** Field effect transistors characterizations after 150°c annealing: a) sfer curve at $V_{GS}$ = 30 V and b) output curves.

### 3.5 Morphology

Atomic force microscopy (AFM) was used on a Tri-BTDPP:PC$_{71}$BM (1:1) blend film prepared using the same conditions as for the one that achieved the best solar cell in order to get information on thin film morphology. As shown in Fig. 4, large grains of 100 nm-200 nm were observed in both height and phase images. Since the phase image resembles the height image, each grain seems uniform in composition. Because the Tri-BTDPP phase is highly crystalline as revealed by the X-ray diffraction (XRD) measurements (to be discussed below), the grains appearing in the AFM images are likely the Tri-BTDPP phase. The PC$_{71}$BM phase might be present at the boundaries of the Tri-BTDPP (the grain boundaries in the phase image are larger than in the height image). It is also possible that a vertical phase separation occurs with a PC$_{71}$BM enriched bottom.[26-28] The increase in J$_{sc}$ observed in the inverted devices compared with the conventional devices supports this hypothesis because the charge collection would be more favored for the inverted structure if PC$_{71}$BM is enriched at the bottom and Tri-BTDPP is enriched on the top. The exciton diffusion length in organic semiconductors is in the range of 1-10 nm;[29,30] the domain size of the Tri-BTDPP phase is therefore too large for efficient exciton diffusion to a donor/acceptor interface. This, in turn, causes poor charge carrier generation due to geminate recombination of excitons and thus a low J$_{sc}$.

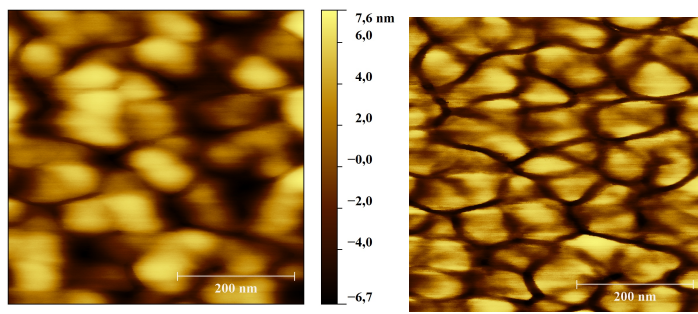

**Table 2** Photovoltaic performances in BHJ solar cells with different architectures

For these reasons, the size of the domains should be reduced to improve the charge separation.

**Fig. 4** AFM heigh (a) and phase (b) images 0.5 µm x 0.5 µm of Tri-BTDPP: PC$_{71}$BM (1:1) film from chloroform and 1%vol DIO solution

To further understand the composition and the origin of the morphology, the interaction of Tri-BTDPP and PC$_{71}$BM was investigated using differential scanning calorimetry (DSC). The measurements were completed on films prepared by drop-casting solutions of Tri-BTDPP and PC$_{71}$BM with different ratios in chloroform. The DSC trace of the first heating scan of each sample is shown in Fig. 6. The melting points of both Tri-BTDPP and PC$_{71}$BM appear independent of the blend ratio, revealing their immiscibility. This indicates that Tri-BTDPP and PC$_{71}$BM do not intermix and rather form respective pure crystalline domains, which explains the observed morphology. According to some studies, an ideal morphology should provide not only pure crystalline domains of donor and acceptor to extract the charges but also donor-acceptor intermixed domains.[31–33] The ultrafast charge separation occurred primarily in intermixed domains. The immiscibility of Tri-BTDPP and PC$_{71}$BM induces the formation of the over-sized pure crystalline domains and prevents the formation of intermixed domains, leading to hampered exciton diffusion and charge separation.

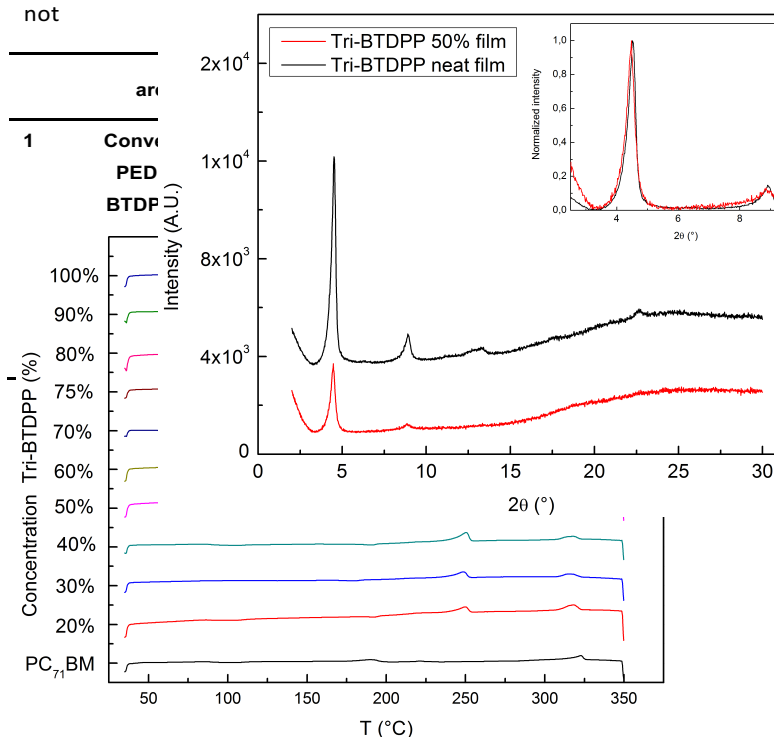

XRD measurements were carried out to study the nature of the molecular packing of the neat Tri-BTDPP and Tri-BTDPP:PC$_{71}$BM films. As shown in Fig. 5, for the Tri-BTDPP neat film, sharp and intense diffraction peaks at 2θ = 4.50°, 8.96°, 13.28°, 17.72° and 22.64° are clearly seen. According to Bragg's law, the first four peaks represent the 1st to 4th order peaks of planes with a *d*-spacing distance of 1.96 nm, while the last peak represents another set of planes with a *d*-spacing distance of 0.39 nm. This XRD pattern is reminiscent of a lamellar packing motif, which has been frequently observed for many crystalline polymers such as regioregular head-to-tail poly(3-hexylthiophene) (P3HT).[34] Therefore it is reasonable to consider that the Tri-BTDPP molecules adopted a lamellar packing motif, with an interlayer distance of 1.96 nm, separated by the alkyl chains, and a π-π stacking distance of 0.39 nm between the conjugated backbones. The size of the crystallites was determined from the primary diffraction peak by the Scherrer equation to be 27.4 nm. For the 1:1 Tri-BTDPP:$PC_{71}BM$ blend film, the primary and secondary diffraction peaks were still observed at the same positions. The size of tri-BTDPP crystallites remained to be 27.4 nm, which confirmed that $PC_{71}BM$ did not intermix with Tri-BTDPP to influence the crystal structure of the Tri-BTDPP phase. To achieve a better morphology, the size of Tri-BTDPP's crystallite may have to be reduced to achieve smaller domains. In some accounts, nucleating agents were used to increase the number of nuclei and to limit the formation of large crystals.[35,36] Alternatively, side chain engineering, *e.g.*, the use of large side chains on the two terminal DPP units, may reduce the crystal size and improve the morphology of the blend films.

**Fig. 5** XRD measurements on Tri-BTDPP neat film and blend with PC$_{71}$BM (1:1)
**Fig. 6** DSC first heatifilms

### 3.6 External quantum efficiency

To further investigate the photovoltaic process within the device, the external quantum efficiency was measured for optimal solar cells. Fig. 7 shows that the current was mainly generated in the range from 300 nm to 600 nm. This spectrum region corresponds to the $PC_{71}BM$ absorption band, indicating that the current is mainly contributed by the excitons formed within $PC_{71}BM$. The lack of tri-BTDPP's exciton contribution can be explained either by the unfavorable morphology that limits charge separation or by the small energy offset between LUMOs of Tri-BTDPP and $PC_{71}BM$ that limits the electron transfer. The morphology had already been shown to be responsible for the limited charge separation due to the oversized Tri-BTDPP grains. The offset between the LUMO of Tri-BTDPP (-3.99 eV) and $PC_{71}BM$ (-4.3 eV)[37] was estimated to be 0.3 eV, which is equivalent to the minimal value needed for electron transfer from donor to acceptor.[22] Therefore, another strategy to improve the cell performance of Tri-BTDPP devices may be to slightly raise the LUMO level of Tri-BTDPP or to use other electron acceptors with lower LUMO levels than that of $PC_{71}BM$.

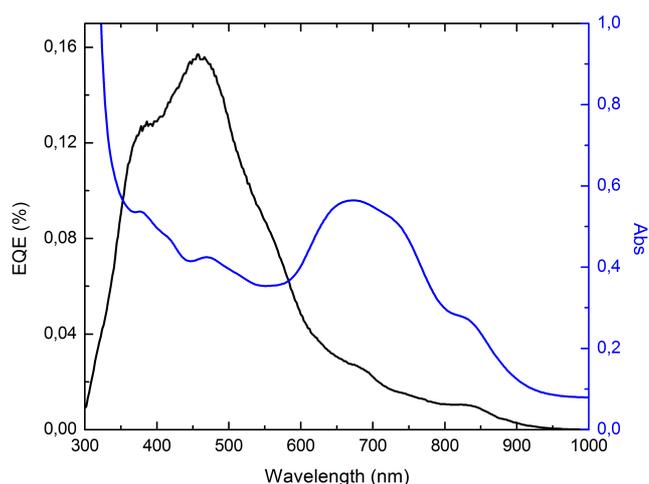

**Fig. 7:** EQE and absorption of Tri-BTDPP/$PC_{71}BM$ blend

## 4. Conclusions

A new oligomer semiconductor containing three bisthiophene-DPP units (Tri-BTDPP) was designed and synthesized via Stille coupling. The extended π-conjugation and strong intersystem crossing transfer of Tri-BTDPP induced a broad absorption up to the near infrared region (~1000 nm). Tri-BTDPP demonstrated a good hole mobility up to 1.22 × $10^{-3}$ cm² $V^{-1}$ $s^{-1}$ in solution-processed OFET. As an electron donor semiconductor in organic solar cells, Tri-BTDPP exhibited a PCE of 0.72%. A careful study of the morphology by AFM and the interactions between Tri-BTDPP and $PC_{71}BM$ by DSC and XRD revealed that Tri-BTDPP and $PC_{71}BM$ were unable to intermix effectively, resulting in oversized Tri-BTDPP crystalline phases and thus poor charge separation. Using processing additives or a nucleating agent may be an effective method to improve the morphology of such bulk heterojunction to produce efficient solar cells.

## Acknowledgements

This work was financially supported by the University of Bordeaux and SOLVAY in framework of IDS-FUNMAT network (2012-14-LF). We would like to acknowledge Bertrand Pavageau, Marie-Béatrice Madec and Bin Sun for their contribution in the work.